# Linear and Non Linear Behavior of Mechanical Resonators for Optimized Inertial Electromagnetic Microgenerators.

C. Serre[a], A. Pérez-Rodríguez[a], N. Fondevilla[a], E. Martincic[c], J.R. Morante[a], J. Montserrat[b], and J. Esteve[b]

[a] EME/CEMIC/CERMAE – Dept. Electrònica, Univ. Barcelona, 08028 Barcelona, Spain
[b] Centre Nacional de Microelectrònica IMB-CSIC, 08193 Bellaterra, Spain
[c] IEF, Université Paris-Sud-XI, 91405 Orsay, France

**Abstract-** In this work, the design, fabrication and characterization of an electromagnetic inertial microgenerator compatible with Si micro-systems technology is presented. The device includes a fixed micromachined coil and a movable magnet mounted on a resonant polymeric structure. The characterization of the fabricated prototypes has allowed to observe the presence of non linear effects that lead to the appearance of hysteretic vibrational phenomenon. These effects are likely related to the mechanical characteristics of the polymeric membrane, and determine an additional dependence of vibration frequency on the excitation amplitude. Under such non linear conditions, power densities up to 40 $\mu W/cm^3$ are obtained for devices working with low level excitation conditions similar to those present in domestic and office environment.

## I. INTRODUCTION

The need for wearable or abandoned microsystems, as well as the trend to a lower power consumption of electronic devices, make miniaturized renewable energy generators a viable alternative to batteries. Among the different alternatives, an interesting option is the use of inertial microgenerators for energy scavenging from vibrations present in the environment. These devices constitute perpetual energy sources without the need for refilling, thus being well suited for abandoned sensors, wireless systems or microsystems which must be embedded within the structure, without outside physical connections. Different electromagnetic energy scavenging devices have been described in the literature [1,2,3], based on the use of a velocity damped resonator, which is well suited for harvesting of vibrational energy induced by the operation of machines. These vibrations are characterized by a well defined frequency (in the range between few kHz's and few kHz's) and low displacement amplitudes. Adjusting the resonant frequency of the system to that of the vibrations allows amplification of these low amplitude displacements. Moreover, for these applications, the use of an electromagnetic device has the potential advantages of a good level of compatibility with Si Microsystem technology, as well as the possibility of relatively high electromechanical coupling with simple designs.

This work describes the design and implementation of an electromagnetic inertial microgenerator prototype for energy scavenging from residual ambient vibrations. It is based on a fixed coil and a movable magnet mounted on a resonant structure, according to the optimized power design previously reported in [8]. For the fabrication of the device, a modular manufacturing process has been designed, in which the electromagnetic converter and the mechanical resonator are micromachined separately, diced and then assembled. In contrast with previous works [7], where conventional wound coils were used, a special effort has been devoted here to the development of micromachined coils. Although this determines a significant reduction of the device output, this allows to take advantage of robustness, low cost, reliability and reproducibility features inherent to the microsystems technology and allow the required degree of integration to achieve fully autonomous self-powered remote devices.

## II. DESIGN OF THE MICROGENERATOR PROTOTYPE

The mechanical resonator is constituted by a square 7x7 mm$^2$ magnet, stuck on a Kapton membrane glued between two pcb frames pressed with screws. This polymer has a Young's module of about 2.5 GPa, which is significantly lower than that of Si related materials. This makes it better suited for resonant applications in the frequency range of few Hz to few kHz. The membrane size in combination with the thickness of the polymer allows to tune the microgenerator to specific applications needs over a broad range of resonant frequency. As shown in fig. 1, the design includes a second magnet below the membrane (holding magnet), which allows fixing the inertial mass without the need to glue it. This has led to a significant improvement in the fabrication process and device operation reproducibility, since the free standing membrane area is now perfectly defined by the magnets and the surrounding frame.

The electromagnetic converter module has been designed with a micromachined fixed coil with a surface of about 1 cm$^2$. The design of the device has been based on a previous FE analysis (ANSYS) of the magnetic flux rate as a function of the device parameters [4, 8]. According to this analysis, the

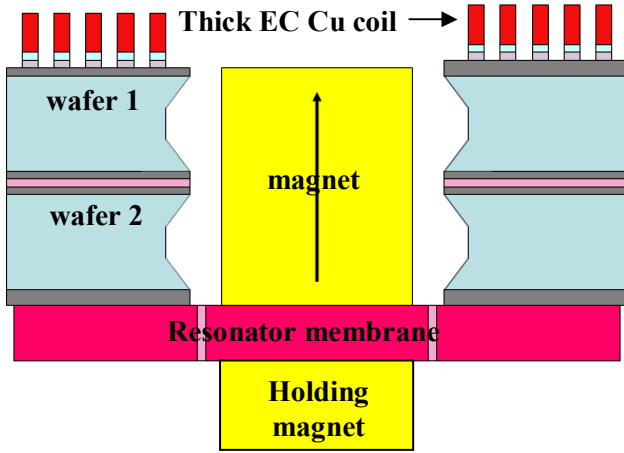

**Thick EC Cu coil** →

wafer 1

wafer 2

magnet

**Resonator membrane**

**Holding magnet**

Figure 1. Electromagnetic inertial microgenerator made of a fixed thick electro-chemically deposited coil (EC coil) and a movable magnet (inertial mass) mounted on a resonant structure.

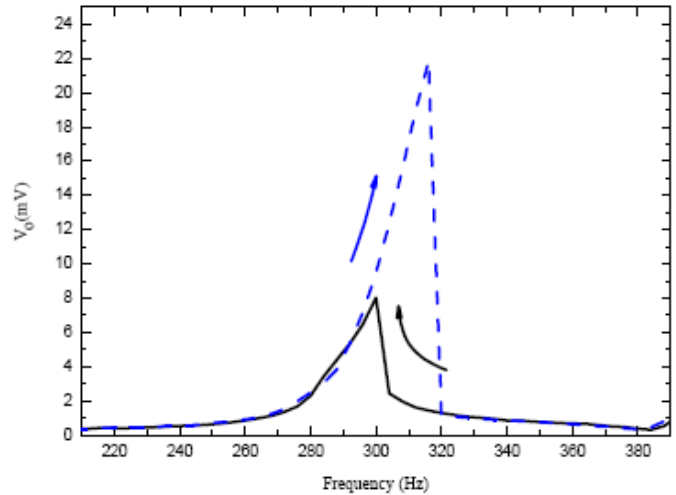

Figure 2. Output voltage $V_0$ versus frequency for a device with a 11x11 mm$^2$ resonator measured under increasing (dotted line) and decreasing (full line) frequency sweep.

electromagnetic coupling can be increased by optimizing the coils geometry in terms of magnetic flux rate, by increasing the track density in the coils while minimizing the series resistance of the coil tracks. Taking this into account, the electromagnetic converter has been fabricated by combining standard microsystems technology processes, with electroplating post-processes for the fabrication of coils with thick Cu metal tracks by electrochemical deposition. This has also involved an optimization of the EC growth conditions to minimize residual stress in the Cu layers and to optimize their adherence, since these are critical parameters that can compromise the viability and lifetime of the devices.

The results of the experimental mechanical and electrical characterization of partially optimized prototypes made of resonators with different geometries, are presented. The effects of stress stiffening on the parasitic damping in the structure in linear and non-linear operation of the resonator strongly affects the output of the microgenerator.

### III. CHARACTERIZATION OF THE PROTOTYPES

Once assembled with the electromagnetic converter, the resulting device is stuck on a piezoelectric stack which provides a well controlled mechanical excitation (up to15 μm displacement at 100 V, with a resonant frequency at 69 kHz, i.e. far from our range of interest), and placed inside a vacuum chamber.

Fig. 2 shows the output voltage measured from a device with a micromachined 52 turns 15 μm thick Cu coil (tracks: 20 μm width and 15 μm separation), on a 11x11 mm$^2$ resonator with a load resistance $R_L = 10^5$ Ω and an excitation amplitude $Y_0 = 0.8$ μm. The curves plotted in this figure correspond to the data obtained from an increasing (dotted line) and decreasing (full line) frequency sweeps. These data show the existence of an hysteresis phenomena, that is related to the presence of non linear effects due to spring stiffening of the polymeric membrane [6]. As shown in this figure, the gradual increase of the excitation frequency determines the presence of a metastable oscillating behavior, obtaining a significant increase in the vibration amplitude and, hence, the output voltage, and the resonant frequency in relation to those achieved under decreasing or constant frequency conditions. This hysteresis enhanced output in non linear mode is also related to a decrease of the apparent parasitic damping in the resonator structure with respect to stationary conditions .

The measurements performed at increasing excitation amplitude $Y_0$, under increasing frequency sweeps are plotted in fig.3, which shows the output voltage measured with $R_L = 105$ Ω from a device with a 15x15 mm$^2$ resonator. As reported in [5], these loading conditions correspond to the optimal ones for which a maximum value of generated power and voltage are achieved at the device terminals. The saturation observed at high excitation amplitude is correlated to an increase of the apparent parasitic damping.

The dependence of the hysteresis enhanced output of the micro-generator on the loading resistance can be seen in Fig. 4 which shows the output power and voltage measured at resonant conditions versus the load resistance, According to these measurements, the fabricated prototypes are able to deliver a maximum output power of the order of 50 μW and a maximum voltage of about 180 mV for excitation conditions corresponding to f = 344 Hz, $Y_0$ = 5.1 μm. Fitting of the experimental data gives a value of the parasitic damping of $\zeta_p$ = 0.008.

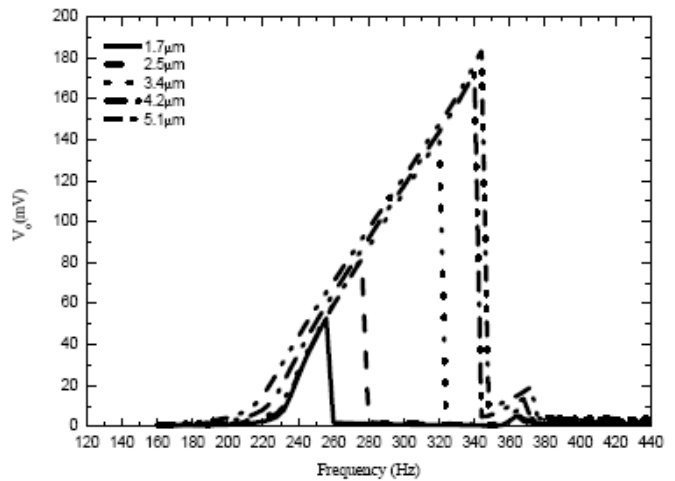

Figure 3. Output voltage $V_0$ versus frequency for a device with a 15x15 mm$^2$ resonator measured under increasing frequency sweep with different excitation amplitudes $Y_0$.

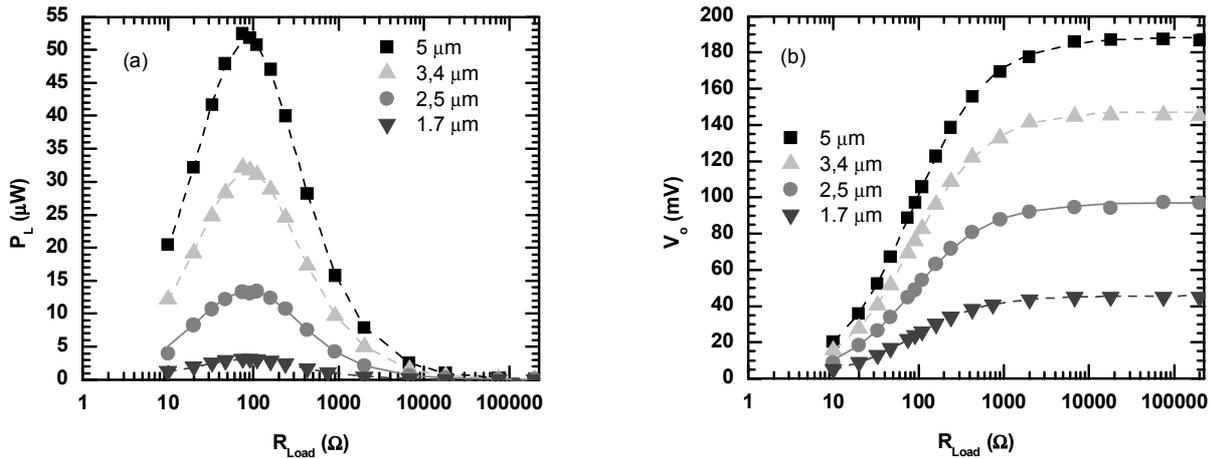

Figure 4. Output power $P_L$ (a) and output voltage $V_o$ (b) vs. load resistance $R_L$ with different excitation amplitudes $Y_0$, for a 15x15 mm$^2$ square Kapton resonator membrane, with a 15 μm thick electroplated Cu coils (52 turns, track width = 15 μm, track separation = 20

Using the active device volume (of the order of 1.35 cm$^3$) and the excitation acceleration, the experimental data give a normalized power density in the range between 1.6 x10$^{-6}$ W/cm$^3$g$^2$ and 3.8x10$^{-6}$ W/cm$^3$g$^2$. These values agree with those expected from the theoretical scaling trend reported by Arnold in [7]. In that work, a dependence of the power density with the fourth power of the linear dimensions is predicted for miniaturized electromagnetic devices. Even if higher values of the power density are reported, it is important to bear in mind that many of these values are achieved with resonators implementing conventional wire windings for the transducer coil. This contrasts with our device, which uses processes fully compatible with Si micromachining technology for the fabrication of the electromagnetic transducer. Moreover, it is interesting to remark that further optimization of the thick resist photolithographic processes would allow increasing the thickness of the EC grown Cu tracks in the coils, which would lead to a further improvement of the generated power. In this sense, increasing the thickness of the metal tracks in the coils to values in the range from 20 to 40 μm would allow increasing of the generated power up to values in the range of hundreds of μW.

However, the metastable behavior of the resonator implies the need to work under increasing excitation frequency conditions to achieve the maximum power values. In addition, the dependence of the peak frequency on the excitation amplitude has also to be taken into account when designing the device for a practical application, where the peak frequency has to be tuned to that of the vibrations present in the ambient.

ACKNOWLEDGMENTS

EME is a member of CEMIC (Center of Microsystems Engineering) of the "Generalitat de Catalunya". The funding of this work by the IST program of the European Commission under the SENSATION project (ref. FP6-507231) is acknowledged by the authors from the University of Barcelona. This work was also supported by grant SAB-2006-165 to E. Martincic from the Spanish "Ministerio de Educación y Ciencia".